\documentstyle[aps,prl,epsf,floats]{revtex}

\begin{document}

\twocolumn[\hsize\textwidth\columnwidth\hsize\csname@twocolumnfalse%
\endcsname

\title{Finite Temperature Depinning of a Flux Line from a Nonuniform 
Columnar Defect}
\author{Deniz Erta\c s}
\address{Exxon Research \& Engineering Company,
Clinton Twp., Rte 22 East,
Annandale, NJ 08801}
\date{\today}
\maketitle

\begin{abstract}
A flux line in a Type-II superconductor with a single nonuniform columnar 
defect is studied by a perturbative diagrammatic expansion around an annealed 
approximation. The system undergoes a finite temperature
depinning transition for the (rather unphysical) on-the-average 
repulsive columnar defect, provided that the fluctuations along the axis are 
sufficiently large to cause some portions of the column to become attractive. 
The perturbative expansion is convergent throughout the weak pinning regime
and becomes exact as the depinning transition is approached, providing an 
exact determination of the depinning temperature and the divergence of the 
localization length. 
\end{abstract}

]

The finite-temperature statistical behavior of magnetic flux lines (FLs) in 
Type-II superconductors with various kinds of disorder has been a topic of 
considerable interest and technological relevance following the advent of 
high-$T_c$ superconductors about a decade ago\cite{Blatter}. Early on, it 
was realized that compared to point defects, extended defects such as twin 
boundaries and columnar amorphous regions created by heavy ion 
irradiation\cite{Civale} are much more effective in pinning of FLs
at higher temperatures, an essential feature for technological 
applications of these new materials. Most analyses of extended defects 
studied uniform structures\cite{Nelson:93}. However, it is natural to expect 
that such defects contain spatial inhomogeneities in themselves.
This paper addresses the implications of such a possibility for the simple 
case of a single FL in the presence of a single inhomogenous columnar defect 
with short-range correlated disorder along its axis, which is aligned with 
the external magnetic field. 

A number of closely related works on 2D wetting with quenched disorder%
\cite{Forgacs,Derrida}, directed polymers with random interaction\cite{Mukherji},
strongly inhomogeneous surface growth\cite{Kallabis}, and
2D wetting and directed polymers in a periodic potential\cite{Nechaev,Galluccio}
have mostly considered a
(1+1) dimensional geometry, for which the disorder was marginally relevant
at the depinning transition. Higher dimensional systems were glossed over 
upon the observation that disorder was irrelevant, i. e., the critical behavior
near the depinning transition was the same as the uniform defect case.
Nevertheless, it would be very useful to compute the change in the free energy
and the localization length due to quenched disorder, in some vicinity of 
the depinning transition where small-scale details of the disorder becomes 
unimportant. This paper presents a controlled perturbation expansion, which 
becomes exact in the weak pinning limit, for the corrections
to the free energy of the FL due to fluctuations in the pinning potential 
along the columnar defect. 

Consider a single FL (with line tension $\tilde{\epsilon}$) in the
presence of a columnar defect of radius $b$, oriented in the $z-$direction
with a pinning potential $U(z)$ inside the defect. 
The Hamiltonian for a FL configuration $\{{\bf r}(z)\}$ is 
\begin{equation}
{\cal H}(\{{\bf r}(z)\})=\int_{0}^{L}{dz\left\{ {\frac{\tilde{\epsilon}}{2}}%
\left( {\frac{d{\bf r}}{dz}}\right) ^{2}+\left[ U_{0}+\delta U(z)\right] h(%
{\bf r}(z))\right\} },
\end{equation}
where $h({\bf r})=\Theta (b-|{\bf r}|)$.($\Theta $ is the unit step function.) 
The fluctuations $\delta U(z)$ around the mean potential $U_0$
have a Gaussian distribution with zero mean and correlations 
\begin{equation}
\overline{\delta U(z)\delta U(z^{\prime })}=\frac{\sigma ^{2}}{2\pi }\exp
\left( -\frac{(z-z^{\prime })^{2}}{2\xi ^{2}}\right) ,
\end{equation}
where the overline represents averaging over the disorder potential. Various
thermodynamic properties of the system such as the free energy can be
deduced from the partition function of the system: 
\begin{equation}
Z=Z_{{\rm free}}^{-1}\int {{\cal D}{\bf r}(z)\exp \left[ -\beta {\cal H}(\{%
{\bf r}(z)\})\right] },
\end{equation}
where $\beta =1/T$ ($k_{{\rm B}}\equiv 1$ throughout), and the normalization factor 
is chosen such that $Z=1$ for $U_{0}=\delta U(z)=0$. $Z$ is a random variable 
that depends on the particular realization of disorder. Thermodynamic quantities 
typically involve averaging of $\ln Z$, which is usually too
difficult to obtain directly. An alternative approach is to calculate
moments of the random variable $Z$ since $\overline{Z^{n}}=\overline{e^{n\ln Z}%
}$ is the characteristic function for the random variable 
$\ln Z$\cite{Kardar:87}. Indeed, 
\begin{equation}
\overline{Z^{n}}=\exp \left[ \sum_{k=1}^{\infty }\frac{n^{k}}{k!}C_{k}(\ln
Z)\right] ,
\end{equation}
where $C_{k}(\ln Z)$ is the $k$th cumulant of $\ln Z$. The $n$th moment 
of $Z$ is 
\begin{eqnarray}
Z^{n}&=&\int {\cal D}{\bf r}_{1}(z)...{\cal D}{\bf r}_{n}(z)\exp \left\{
-\beta \int dz\sum_{i=1}^{n}\frac{\tilde{\epsilon}}{2}\left( \frac{d{\bf r}%
_{i}}{dz}\right) ^{2} \right. \nonumber \\
& &\qquad \left.+[U_{0}+\delta U(z)]\sum_{i=1}^{n}h({\bf r}_{i}(z))%
\right\} .
\end{eqnarray}
Upon averaging over disorder, one obtains expectation values of the form 
\begin{eqnarray}
\overline{e^{-\beta \int dz\,m(z)\delta U(z)}} &=&e^{\frac{\beta
^{2}}{2}\int dz dz^{\prime }\,m(z)m(z^{\prime })\overline{\delta
U(z)\delta U(z^{\prime })}}  \nonumber \\
&\approx &\exp \left\{ \frac{\beta ^{2}\sigma ^{2}\xi}{2}\int dz\,[m(z)]^{2}%
\right\}
\end{eqnarray}
where $m(z)\equiv \sum_{i=1}^{n}h({\bf r}_{i}(z))$ is the number of
lines that are inside the defect at height $z$, and the
approximation $m(z^{\prime })\approx m(z)$ leading to the final result is
good provided that $\xi < \xi _{z}\equiv\beta\tilde{\epsilon}b^{2}$, the 
diffusion length inside the defect. Using $%
[m(z)]^{2}=m(z)+m(z)[m(z)-1]$, the $n$th moment can be rewritten as 
\begin{eqnarray}
\overline{Z^{n}}&=&\int \prod_{i=1}^{n}\left\{ {\cal D}{\bf r}_{i}(z)\exp
\left[ -\beta {\cal H}_{0}(\{{\bf r}_{i}(z)\})\right] \right\} \nonumber \\
& & \quad \times \exp \left[
\beta ^{2}\sigma ^{2}\xi \int dz\sum_{i<j}h({\bf r} _{i}(z))h({\bf r}%
_{j}(z))\right] ,  \label{eqZn}
\end{eqnarray}
where 
\begin{equation}
{\cal H}_{0}(\{{\bf r}_{i}(z)\})=\int_{0}^{L}{dz\left\{ {\frac{\tilde{
\epsilon}}{2}}\left( {\frac{d{\bf r}}{dz}}\right) ^{2}+ U_{{\rm eff}} 
h({\bf r}(z))\right\} }
\end{equation}
is the Hamiltonian corresponding to a FL in the presence of a {\it uniform}
columnar defect, with defect energy per unit length 
\begin{equation}
U_{{\rm eff}}(T)=U_{0}-\sigma ^{2}\xi /2 T.
\end{equation}

The solution to the uniform cylindrical defect problem is well 
studied\cite{Nelson:93}. The free energy per unit length $f_0$ and probability 
density $\psi _{0}^{2}({\bf r})$ of the FL at a position ${\bf r}$ are given 
respectively by the ground state energy and wavefunction of the corresponding
two-dimensional ``Schr\"{o}dinger Equation'' 
\begin{equation}
\left[ -\frac{T^{2}}{2\tilde{\epsilon}}\nabla _{\perp }^{2}+U_{%
{\rm eff}}h({\bf r})\right] \psi _{0}({\bf r})=f_{0}\psi _{0}({\bf r}).
\end{equation}
For $U_{{\rm eff}}<0$, the FL is pinned to the defect, with 
\begin{eqnarray}
f_{0} &\approx &-\frac{\left| U_{{\rm eff}}\right| }{2}\left( \frac{T}{
T_{\ell }}\right) ^{2}e^{-2(T/T_{\ell })^{2}},\;T\gg T_{\ell }, \\
\ell _{\perp } &\equiv &\left[ \int d{\bf r}\,r^{2}\psi _{0}^{2}({\bf r}%
)\right] ^{1/2}\approx be^{(T/T_{\ell })^{2}},\;T\gg T_{\ell },
\end{eqnarray}
where $\ell _{\perp }$ is the localization length and 
$T_{\ell }\equiv \sqrt{-U_{{\rm eff}}\tilde{\epsilon}}
\,b $ is the localization temperature, below which the FL is strongly pinned to
the defect $(\ell _{\perp }\approx b)$ and the free energy is dominated by
the ground state energy rather than the entropy $(f_{0}\approx U_{{\rm eff}%
}) $. 

For $U_{0}<0$ (on-average attractive defect), $U_{{\rm eff}}<0$ and the FL 
is pinned to the defect at all temperatures. However, for $U_{0}>0$ (on-average 
repulsive defect), there is a depinning transition when $U_{{\rm eff}}=0$, at
a temperature 
\begin{equation}
T_{d}=\frac{\sigma ^{2}\xi }{2 U_{0}}.
\end{equation}
For $T>T_{d}$, the FL is delocalized, with $f_{0}=0$. 
As $t\equiv (T_{d}-T)/T_{d} \searrow 0^+$, the free energy and localization 
length diverge as 
\begin{eqnarray}
f_{0}(t) &\approx &-\frac{U_{0}}{2}\frac{\sigma ^{4}}{t\sigma _{c}^{4}}\exp
\left[ -{\frac{2}{t}}\frac{\sigma ^{4}}{\sigma _{c}^{4}}\right] ,
\label{eqf0} \\
\ell _{\perp }(t) &\approx &b\exp \left[ {\frac{1}{t}}\frac{\sigma ^{4}}{
\sigma _{c}^{4}}\right] , \label{eqlperp}
\end{eqnarray}
where $\sigma _{c}\equiv U_{0}^{3/4}\tilde{\epsilon}^{1/4}(2b/\xi )^{1/2}$.

Notice that 
\begin{equation}
\label{eqZn2}
\overline{Z^{n}}=Z_{0}^{n}\left\langle \exp \left[ \beta ^{2}\sigma ^{2}\xi
\int dz\sum_{i<j}h({\bf r}_{i}(z))h({\bf r}_{j}(z))\right] \right\rangle
_{0},
\end{equation}
where $Z_{0}=e^{-\beta f_{0}L}$ and $\langle ...\rangle _{0}$ denotes an
average over a canonical ensemble of $n$ {\it non-interacting} FLs in the
presence a uniform defect of strength $U_{{\rm eff}}$. 
This additional exponential factor is due to the fact that all $n$ FLs
sample the same ``quenched'' realization of disorder, and can be interpreted 
as an effective two-body interaction between FLs that is nonzero {\it only when
both FLs are inside the defect}. 

\begin{figure}[tbp]
\narrowtext
\epsfxsize=2.9truein
\vbox{\hskip 0.15truein
\epsffile{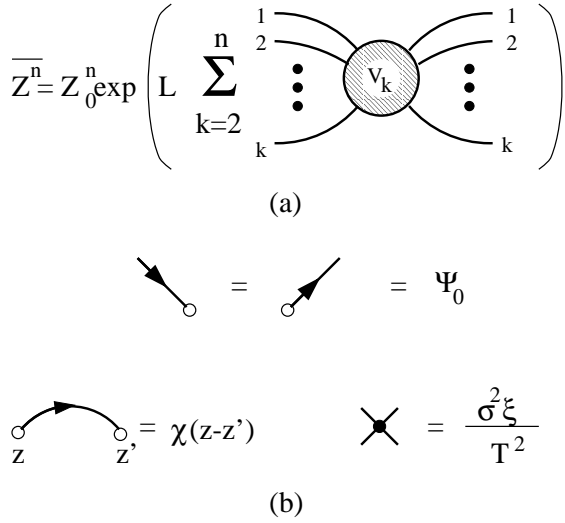}}
\medskip
\caption{(a) Diagrammatic representation of the disorder-averaged partition
function. (b) Feynman rules for calculating various diagrams that contribute
to vertex functions $V_k$.}
\label{feynmanrules}
\end{figure}

For $T>T_b$, the FLs are delocalized and the free energy for the quenched 
system is identically 0.
In the weak pinning regime $T_d > T \gg T_\ell$, corrections to the free energy 
 can be computed as a controlled perturbation series in some ``small'' parameter.
Expanding Eq.(\ref{eqZn2}) in a power series: 
\begin{eqnarray}
\frac{\overline{Z^{n}}}{Z_0^n}&=&\sum_{k=0}^{\infty }\frac{(\beta
^{2}\sigma ^{2}\xi )^{k}}{k!}\sum_{i_{1}<j_{1}}...\sum_{i_{k}<j_{k}}\int
dz_{1}...dz_{k}  \\
& &\quad\times\left\langle h({\bf r}_{i_{1}}(z_{1}))h({\bf r}%
_{j_{1}}(z_{1}))...h({\bf r}_{i_{k}}(z_{k}))h({\bf r}_{j_{k}}(z_{k}))\right%
\rangle _{0}. \nonumber
\end{eqnarray}
The expectation value factors for distinct FLs since each FL wanders
independently of each other. After ordering in $z$ such that $z_i\leq z_j$
for $i<j$, the product can be further factorized as 
\begin{eqnarray}
\left\langle \prod_{i=1}^{s}h({\bf r}(z_{i}))\right\rangle _{0}&=&\Psi
_{0}^{2s}\prod_{i=1}^{s-1}\frac{\left\langle h({\bf r}(z_{i}))h({\bf r}
(z_{i+1}))\right\rangle _{0}}{\Psi _{0}^{4}}, \\
\Psi _{0}^{2}\equiv \left\langle h({\bf r}(z))\right\rangle
_{0}&=&\int\limits_{|{\bf r}|<b}d{\bf r}\,\psi _{0}^{2}({\bf r}) \nonumber \\
&\approx& \pi
\left( \frac{T}{T_{\ell }}\right) ^{4}\,e^{-2(T/T_{\ell })^{2}},
\end{eqnarray}
since each visit to the defect (at $z=z_i$) is a renewal event, i. e., 
the configuration of the FL for $z>z_i$ is independent of its configuration 
for $z<z_i$. However, the probability of a subsequent visit to the defect
is enhanced up to a distance of order 
$\ell _{z}=\xi_{z}(\ell _{\perp }/b)^{2}$, after which the FL ``forgets'' 
its previous visit. This enhancement can be captured in an effective 
``propagator'' 
\begin{equation}
\chi (z)\equiv \frac{\left\langle h({\bf r}(0))h({\bf r}(z))\right\rangle
_{0}-\left\langle h({\bf r}(0))\right\rangle _{0}^{2}}{\left\langle h({\bf r}
(0))\right\rangle _{0}^{2}},
\end{equation}
which contains all the $z-$dependence and becomes exponentially small 
for $z > \ell _{z}$. Then, 
\begin{equation}
\left\langle \prod_{i=1}^{s}h({\bf r}(z_{i}))\right\rangle _{0}=\Psi
_{0}^{2s}\prod_{i=1}^{s-1}[1+\chi (z_{i+1}-z_{i})].
\end{equation}
Upon further manipulation, the disorder-averaged partition function can be
recast into a form [See Fig.~\ref{feynmanrules}(a)] 
\begin{equation}
\overline{Z^{n}}=e^{-nf_{0}L/T}\exp \left[ \frac{n(n-1)}{2}%
L\sum_{k=2}^{\infty }V_{k}\right] ,
\end{equation}
where the vertex functions $V_{k}$ are a sum over all connected graphs with $%
k$ incoming and $k$ outgoing lines, constructed from the building blocks
shown in Fig.~\ref{feynmanrules}(b). The combinatorial factor coming from the 
choice of the first pair to be contracted is pulled out for notational 
convenience.

\begin{figure}[tbp]
\narrowtext
\epsfxsize=2.9truein
\vbox{\hskip 0.15truein
\epsffile{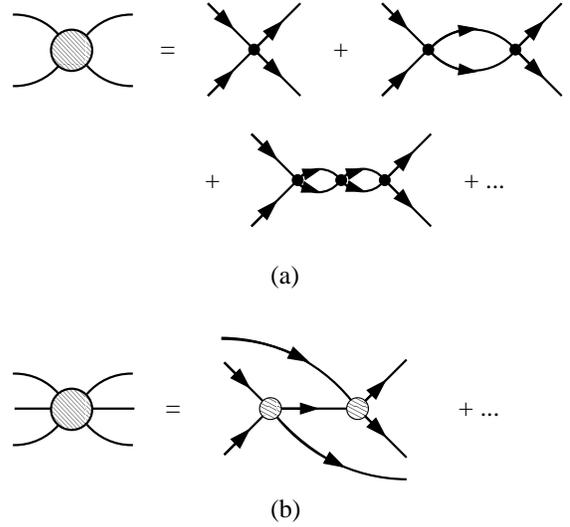}}
\medskip
\caption{(a) Diagrammatic representation of the geometric series that
contributes to $V_2$. (b) Leading-order contribution to $V_3$.}
\label{series}
\end{figure}

Leading-order contributions to the free energy come from $V_{2}$, which can
be computed exactly in terms of $\Psi _{0}$ and $\chi $: As shown in Fig.~%
\ref{series}(a), the terms from a geometric series, and 
\begin{equation}
V_{2}=\frac{\sigma ^{2}\xi }{T^{2}}\Psi
_{0}^{4}\sum_{i=0}^{\infty }\left[ \frac{\sigma ^{2}\xi 
}{T^{2}}\Psi_{0}^{2}\int_{0}^{\infty }dz\,\chi^{2}(z)\right]^{i}.
\end{equation}

Just below the pinning temperature, $0<t\ll 1$, $\Psi
_{0}^{2}\int_{0}^{\infty }dz\chi ^{2}(z)\approx C\xi _{z}$ , where $C$ is a
constant of ${\cal O}(1)$. In this weak pinning limit, the correction to the
disorder-averaged free energy and its variance are given by the
coefficients of terms in the exponent of $\overline{Z^n}$ that are linear 
and quadratic in $n$, respectively. 
\begin{eqnarray}
\overline{\delta f_{{\rm quenched}}} &\approx &\frac{\sigma ^{2}\xi }{2T_{d}}
\frac{\Psi _{0}^{4}}{1-(C\sigma ^{2}\xi \xi _{z}/T^{2})} 
\nonumber \\
&\approx &\pi ^{2}U_{0}\left( \frac{\sigma ^{4}}{\sigma _{c}^{4}t}\right)
^{4}\frac{\sigma ^{4}}{\sigma ^{4}-2C\sigma _{c}^{4}}\exp \left[ -\frac{4}{t}%
\frac{\sigma ^{4}}{\sigma _{c}^{4}}\right] , \\
\overline{\delta f_{{\rm quenched}}^{2}} &=&\frac{ T_{d}}{L}%
\overline{\delta f_{{\rm quenched}}}.
\end{eqnarray}
Thus, the free energy is self averaging as $L\to \infty $. Note that
the geometric series converges when $\sigma^2 > \sqrt{2C}\sigma_{c}^2$. The
leading-order contribution to $V_{3}$  is [See Fig.~\ref
{series}(b)]
\begin{equation}
V_{3}^{(1)}=(n-1)V_{2}^{2}\int_{0}^{\infty }dz\,\chi (z).
\end{equation}

Near the transition, $\int_{0}^{\infty }dz\,\chi (z)\approx (T/T_{\ell
})^{4}/\Psi _{0}^{2}$ and therefore this next correction to $\delta f_{{\rm %
quenched}}$ is smaller by a factor of order $(\sigma ^{4}/\sigma
_{c}^{4}t)^{4}\exp (-2\sigma ^{4}/t\sigma _{c}^{4})$, in fact, leading order
contributions from higher order vertex functions form a series where each
term is smaller from the previous one by this same factor. Thus, the
asymptotic behavior of the free energy and the localization length near the
depinning transition is given by Eqs.(\ref{eqf0}) and (\ref{eqlperp}), since 
\begin{equation}
\lim_{T\to T_{d}^{-}}\frac{\overline{\delta f_{{\rm quenched}}}}{f_{0}}=0.
\end{equation}

The perturbation series can be used to compute corrections to the free
energy even away from the depinning transition, provided that 
$T\gg T_{\ell }$, i. e., within the weak pinning regime. The situation
is clear for $U_0<0$. In the interesting case $U_0>0$ and
$\sigma \gg \sigma _{c}$, this condition breaks down near a temperature 
$\tilde{T}\approx \left(\sigma _{c}^{2}/2\sigma ^{2}\right)^{2/3}T_{d}$.
For $T<\tilde{T}$, the localization length becomes of order $b$ and 
energetic contributions dominate over entropic contributions to the 
free energy. At $T=0$, the FL sits adjacent to the defect and makes
excursions into it whenever energy fluctuations are favorable. The energy
(per unit length) cost of a typical excursion of length $\ell $ is
\begin{equation}
E(\ell )/\ell \approx \tilde{\epsilon}b^{2}/\ell ^{2}+U_{0}-\sqrt{\sigma
^{2}\xi /\ell },
\end{equation}
which is minimized for $\ell ^{*}\approx (16\tilde{\epsilon}^{2}b^{4}/\sigma
^{2}\xi )^{1/3}$ , and yields an estimate for the ground state energy 
\begin{equation}
f(T=0)\approx -\frac{U_{0}}{2}\left[ \frac{3}{4}\left( \frac{2\sigma ^{2}}{%
\sigma _{c}^{2}}\right) ^{2/3}-1\right].
\end{equation}
For $\sigma \lesssim \sigma_{c}$, fluctuations are not favorable enough 
to make excursions into the defect, and the FL is depinned at all temperatures,
which suggests why the perturbation expansion breaks down for this case.

\begin{figure}[tbp]
\narrowtext
\epsfxsize=2.9truein
\vbox{\hskip 0.15truein
\epsffile{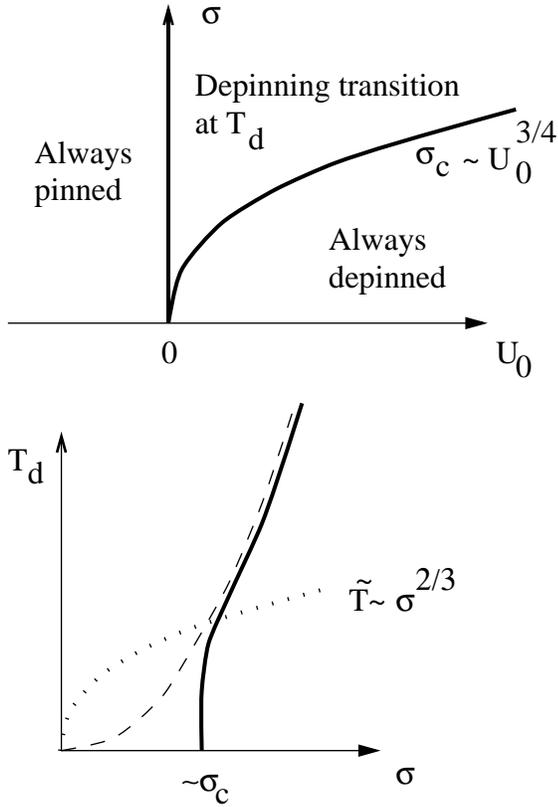}}
\medskip
\caption{{\it Top\ }: Diagram depicting three possible scenarios as a
function of $\sigma$ and $U_0$. {\it Bottom\ }: Depinning temperature $T_b $
as a function of disorder, for a ``repulsive'' defect with $U_0>0$. The
diagrammatic series gives a good approximation to the pinning free energy
whenever $T\gg \tilde T$.}
\label{phasediagram}
\end{figure}

The $T=0$ behavior can be expected persist to finite temperatures as long as 
$\ell^* < \xi_z$, or equivalently, $T \gtrsim \tilde T$. Thus, as 
temperature is increased, there is a simple crossover from strong to 
weak pinning near $\tilde T$. Fig.~\ref{phasediagram} depicts a cartoon 
phase diagram and the weak pinning regime where the perturbation expansion
 is useful.

Finally, the same methodology can be simply generalized to a ($d+1$)-dimensional 
geometry. The perturbation series diverges for $d=1$ as amply demonstrated
by various authors\cite{Forgacs,Derrida,Kallabis}, but useful results for 
all $d>1$ can be obtained. In particular, results for the periodic 
potential\cite{Galluccio} can be readily extended to the random case.  

I would like to thank Professor David R. Nelson for suggesting the problem
and subsequent encouragement. The bulk of this work was completed at 
Harvard University and supported by the NSF through Grant Nos. DMR-9714725, 
9416910, and 9106237.

\end{document}